\documentclass[amsmath,amssymb,reprint,aps,prl,floatfix,superscriptaddress]{revtex4-1}
\usepackage{bm}
\usepackage{amsmath}
\usepackage{graphicx}
\usepackage{braket}
\usepackage{bbm}
\usepackage[hidelinks]{hyperref}

\begin{document}

\title{Intensity interferometry for holography with quantum and classical light}

\author{G.S. Thekkadath}
\email{guillaume.thekkadath@nrc.ca}
\affiliation{National Research Council of Canada, 100 Sussex Drive, Ottawa, K1A 0R6, Canada}
\affiliation{Department of Physics, Imperial College London, Prince Consort Rd, London SW7 2AZ, UK}

\author{D. England}
\affiliation{National Research Council of Canada, 100 Sussex Drive, Ottawa, K1A 0R6, Canada}

\author{F. Bouchard}
\affiliation{National Research Council of Canada, 100 Sussex Drive, Ottawa, K1A 0R6, Canada}

\author{Y. Zhang}
\affiliation{Department of Physics, University of Ottawa, Ottawa, K1N 6N5, Canada}
\affiliation{National Research Council of Canada, 100 Sussex Drive, Ottawa, K1A 0R6, Canada}

\author{M.S. Kim}
\affiliation{Department of Physics, Imperial College London, Prince Consort Rd, London SW7 2AZ, UK}

\author{B. Sussman}
\affiliation{National Research Council of Canada, 100 Sussex Drive, Ottawa, K1A 0R6, Canada}
\affiliation{Department of Physics, University of Ottawa, Ottawa, K1N 6N5, Canada}

\begin{abstract}
As first demonstrated by Hanbury Brown and Twiss, it is possible to observe interference between independent light sources by measuring correlations in their intensities rather than their amplitudes.
In this work, we apply this concept of intensity interferometry to holography.
We combine a signal beam with a reference and measure their intensity cross-correlations using a time-tagging single-photon camera.
These correlations reveal an interference pattern from which we reconstruct the signal wavefront in both intensity and phase.
We demonstrate the principle with classical and quantum light, including a single photon.
Since the signal and reference do not need to be phase-stable nor from the same light source, this technique can be used to generate holograms of self-luminous or remote objects using a local reference, thus opening the door to new holography applications.
\end{abstract}

\maketitle

\section*{Introduction}
Holography is an established technique to record and reconstruct a wave pattern.
It is commonly used in optics and microscopy to extract phase information from an object~\cite{schnars2015digital,park2018quantitative}.
Conventionally, a hologram is prepared by splitting a light source into two beams: one is used to illuminate the object while the other is used as a reference.
The hologram is then obtained by recording the first-order (i.e. amplitude) interference pattern between the two beams.
These must be correlated in phase during the exposure time, which is typically achieved by placing the object inside a phase-stable interferometer.
Thus, it is generally not possible to prepare holograms of self-luminous or remote objects using this approach.

Several recent works have demonstrated that quantum-correlated light can overcome certain limitations in conventional holography.
Refs.~\cite{topfer2022quantum,fuenzalida2023experimental} used frequency-correlated beams to illuminate the object with a different wavelength than the detected light.
Similarly, Ref.~\cite{chrapkiewicz2016hologram} showed that Hong-Ou-Mandel interference between two single photons enables holography without needing phase stability.
Unfortunately, these and other related techniques~\cite{defienne2021polarization} used inherently faint photon pair sources of light which limits their applicability.

Quantum techniques can inspire researchers to develop classical ones achieving similar results, such as with ghost imaging~\cite{bennink2002twophoton}.
In particular, the Hong-Ou-Mandel effect used in Ref.~\cite{chrapkiewicz2016hologram} is a manifestation of a more general second-order interference phenomenon in which one measures field intensity correlations~\cite{mandel1983photon,ou1988quantum}.
Starting with the pioneering work of Hanbury Brown and Twiss~\cite{brown1956test}, intensity interferometry has been demonstrated with a variety of classical and nonclassical light sources~\cite{magyar1963interference, pfleegor1967interference,hong1987measurement,ou1989fourth,pittman2003violation,hessmo2004experimental,rarity2005non,li2008observation,bennett2009interference,kim2013conditions,liu2015second,deng2019quantum,thekkadath2022measuring}.
To observe interference, intensity correlations must be measured within a sufficiently small time window, which typically requires using fast single-pixel photodetectors and coincidence circuits~\cite{ou1988quantum}.
However, thanks to recent advances in camera technology, it is now possible to measure intensity correlations at the single-photon level in a spatially-resolved manner.
While this has mainly been demonstrated with photon-pair sources~\cite{parniak2018beating,ibarra2020experimental,devaux2020imaging,zhang2021high,camphausen2021quantum,ndagano2022quantum,gao2022high,defienne2022pixel,zia2023interferometric}, it also facilitates quantum-inspired imaging techniques that are applicable to classical light~\cite{bache2006coherent, agafonov2008high, oppel2012superresolving,takeda2014spatial,dangelo2016correlation,tamma2016multipath,ihn2017SecondOrder,murray2022two,yung2022jones}, and has led to a renewed interest in using intensity correlations for astronomy~\cite{gottesman2012longer,Stankus2022twophoton,ou2022unbalanced,brown2022interferometric}.

\begin{figure}
\centering
\includegraphics[width=0.65\columnwidth]{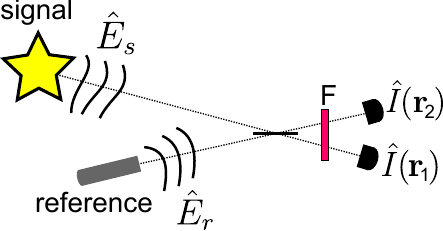}
\caption{\textbf{Intensity interferometry holography}.
The signal (i.e. illuminating or illuminated object) is combined with a reference on a balanced beam splitter.
Unlike in conventional holography, we do not assume that these two fields are phase-stable.
They can be generated by independent sources.
The output intensities are measured using a time-tagging camera.
An interference pattern in the intensity correlations $\braket{\hat{I}(\bm{r}_1)\hat{I}(\bm{r}_2)}$ is observed when the two fields overlap in polarization and time-frequency (which can generally be achieved by a filter $F$ for stationary fields).
This interference pattern depends on the first-order and second-order coherence of the fields [Eq.~\ref{eqn:intensity_cross_correlation}].
}
\label{fig:concept}
\end{figure}

In this work, we present a holography technique based on intensity interferometry.
We combine a signal with a reference and measure their intensity cross-correlations using a time-tagging single-photon camera having nanosecond time resolution.
These correlations reveal a hologram from which we can recover the wavefront of the signal.
Our technique presents two main advantages compared to conventional holography:
(i) it does not require phase stability, and (ii) the signal and reference beams can originate from different sources of light.
To demonstrate these advantages, we perform a proof-of-principle experiment in which we measure the phase transformation of a spatial light modulator.
Firstly, we show that we can record holograms using detector integration times which far exceed the coherence time of our setup.
Secondly, we generate holograms of a single photon and thermal signal using an attenuated laser as the reference beam.
Our technique enables novel applications in holography, such as performing phase imaging with faint light but without the need for phase stabilization, or generating holograms of self-luminous objects using an independent reference.

\section*{Results}
\subsection{Concept}
We begin by describing the principle of the scheme [Fig.~\ref{fig:concept}].
The signal $\hat{E}_s \exp{[i(\phi_s(\bm{r}) + \theta_s)]}$ and reference $\hat{E}_r \exp{[i(\phi_r(\bm{r}) + \theta_r)]}$ fields are combined on a balanced beam splitter.
We assume these fields are co-polarized and have identical time-frequency properties.
Furthermore, the fields must have a fixed locally-varying phase relation but not a global one, i.e. $\phi(\bm{r}) = \phi_s(\bm{r})-\phi_r(\bm{r})$ is fixed while $\theta = \theta_s - \theta_r$ is random.
This arises when the fields are in a quantum state with an indeterminate global phase, such as a single photon or a thermal state.
Alternatively, $\theta$ could be drifting randomly over the course of the measurement due to an unstable interferometer or if the fields are produced by independent light sources.
In either case, if we simply measured the average intensity at a single output port of the beam splitter, i.e. $\braket{\hat{I}(\bm{r}_1)}$, we would not observe any interference.
However, with a temporally-resolving camera, we can instead measure intensity correlations, $G(\bm{r}_1, \bm{r}_2)=\braket{\hat{I}(\bm{r}_1)\hat{I}(\bm{r}_2)}$.
This quantity is determined by pairs of pixels which detected light in both output beams within a correlation window whose duration depends on the light coherence time.
In the Methods, we show that
\begin{equation}
\begin{split}
\tilde{G}(\bm{r}_1, \bm{r}_2)
= \frac{1}{4}&( 2 + \frac{1}{\epsilon} g_s^{(2)}(\bm{r}_1,\bm{r}_2) + \epsilon g_r^{(2)}(\bm{r}_1,\bm{r}_2) \\
& - 2\mathrm{Re}[g_s^{(1)}(\bm{r}_1,\bm{r}_2)g_r^{(1)}(\bm{r}_1,\bm{r}_2)^*] ),
\label{eqn:intensity_cross_correlation}
\end{split}
\end{equation}
where $*$ denotes the complex conjugate and $\sim$ denotes a normalization by the average input intensities whose imbalance is given by $\epsilon=\braket{\hat{I}_{r}}/\braket{\hat{I}_{s}}$.
Eq.~\ref{eqn:intensity_cross_correlation} shows the role of the input field coherence properties in intensity interferometry.
The second-order coherence, $g^{(2)}(\bm{r}_1,\bm{r}_2)$ describes the field correlations only in intensity.
In particular, $g_{s}^{(2)}(\bm{r}_1,\bm{r}_2)$ of the signal could be measured without the reference field, as was done by Hanbury Brown and Twiss~\cite{brown1956test}.
However, by including a reference, Eq.~\ref{eqn:intensity_cross_correlation} contains an additional term which depends on the product of the signal and reference first-order coherence functions, $g^{(1)}(\bm{r}_1,\bm{r}_2)$.
This quantity describes the field correlations in amplitude and phase.
As a result, the last term in Eq.~\ref{eqn:intensity_cross_correlation} reveals an interference pattern which can be used for holographic retrieval of the signal, despite the randomly varying global phase difference $\theta$.
To make this more explicit, we suppose the signal and reference are both in single spatial modes such that $|g_{s,r}^{(1)}(\bm{r}_1,\bm{r}_2)|=1$.
Since these differ by a fixed local phase $\phi(\bm{r})$, we find that
\begin{equation}
\tilde{G}(\bm{r}_1, \bm{r}_2) = \frac{1}{4}\left(A - 2 \cos{[\phi(\bm{r}_1)-\phi(\bm{r}_2)]} \right).
\label{eqn:intensity_corr_purestate}
\end{equation}
Here $A =  2 + g_s^{(2)}/\epsilon + \epsilon g_r^{(2)}$ depends only on the second-order coherence of the inputs and determines the visibility of the interference pattern, $V_0=2/A$.
This equation recovers the familiar upper limit of $V_0=1/2$ ($V_0=1$) for classical (nonclassical) light with Poissonian (sub-Poissonian) photon statistics.
In practice, distinguishability between the signal and reference in time-frequency or polarization will reduce the visibility from these ideal limits~\cite{ou1988quantum}.
We model such imperfections by introducing a phenomenological parameter $M$ such that $V=MV_0=2M/A$ (see Supplementary Materials for more information).
Although a small $V$ reduces the signal-to-noise in the ability to retrieve phase information, we can employ holographic techniques to effectively filter out the noise~\cite{yamaguchi1993phase,schnars1994direct,vargas2011phase}, as we demonstrate experimentally below.

\begin{figure*}
\centering
\includegraphics[width=1\linewidth]{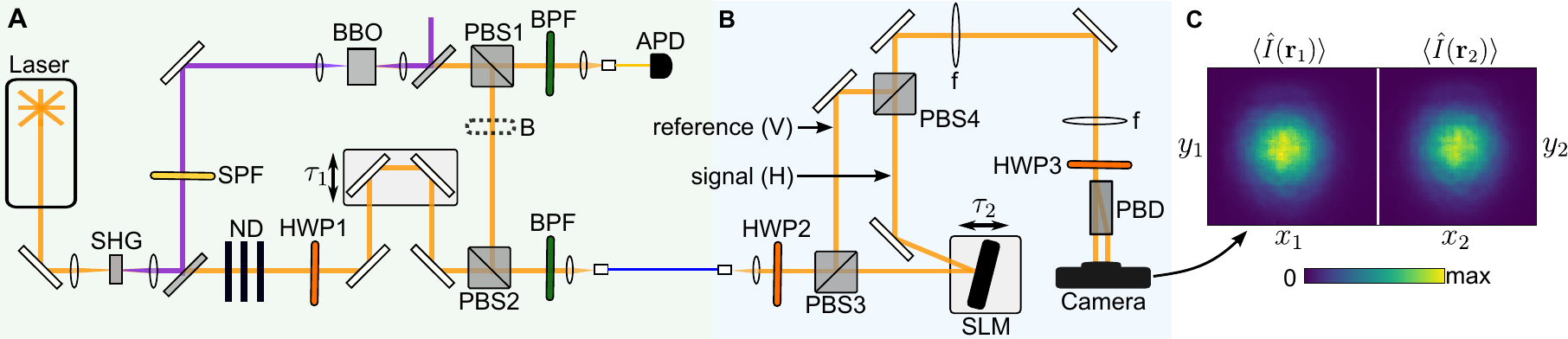}
\caption{\textbf{Experimental setup}.
(A) Signal and reference preparation.
(B) Interferometer and detection.
SHG: second-harmonic generation, SPF: shortpass filter, BPF: bandpass filter, ND: neutral-density filter, BBO: beta barium borate, PBS: polarizing beam splitter, B: beam block, APD: avalanche photodiode, HWP: half-wave plate, SLM: spatial light modulator, H/V: horizontal/vertical polarization, f: convex lens, PBD: polarizing beam displacer.
(C) Typical intensity image of two output beams.
First-order interference is not visible since the beams are phase-uncorrelated.
A 4f lens system images the SLM plane onto the camera.
}
\label{fig:setup}
\end{figure*}

\subsection{Experimental setup}
Our experimental setup [Fig.~\ref{fig:setup}A] allows us to change the signal between either a (i) heralded single-photon state, (ii) phase-randomized coherent state, (iii) thermal state.
The reference is always a coherent state.
Further details are given in the Methods.
The orthogonally-polarized signal and reference are coupled into a polarization-maintaining single-mode fiber and sent toward the interferometer part of the experiment [Fig.~\ref{fig:setup}B]. 
The two beams are separated at PBS3, and the signal is sent onto a spatial light modulator (SLM) which imparts a programmable spatial phase distribution $\phi(\bm{r})$.
The signal is then spatially recombined with the reference at PBS4.
For the sake of demonstration, we purposefully reduce the coherence time of the interferometer consisting of PBS3 and PBS4 to $\sim 100~\mu$s by mounting the SLM onto a vibrating piezo-stage.
To interfere the orthogonally-polarized beams, we rotate their polarization by $\pi/4$ using HWP3 and finally separate the horizontal and vertical components into two beams using a polarizing beam displacer.
We detect the two beams using a single-photon time-tagging camera having 8.3(6) ns timing resolution and 7(2)\% efficiency~\cite{vidyapin2023characterisation}.
The camera sensor is divided into two regions (60 pixels by 60 pixels) corresponding to the two output beams [Fig.~\ref{fig:setup}C].
Each detection event is assigned a label: $\bm{r}_{1i}=(x_{1i}, y_{1i})$ and $t_{1i}$ if the $i$th photon is detected in the left region, and otherwise $\bm{r}_{2i}=(x_{2i}, y_{2i})$ and $ t_{2i}$.
The intensity correlation function $G(\bm{r}_1, \bm{r}_2)=\braket{\hat{I}(\bm{r}_1)\hat{I}(\bm{r}_2)}$ is obtained directly (without any normalization or background subtraction) by making a histogram of events where a pixel from each region fired within a time window $\tau_w$.
In the case of the heralded single photon, we also record the time-tags of the APD detector using the camera's time-to-digital converter and include the condition of detecting a herald photon within $\tau_w$, i.e. measure threefold coincidences [see Fig.~\ref{fig:SNR}].

\begin{figure}
\centering
\includegraphics[width=1\columnwidth]{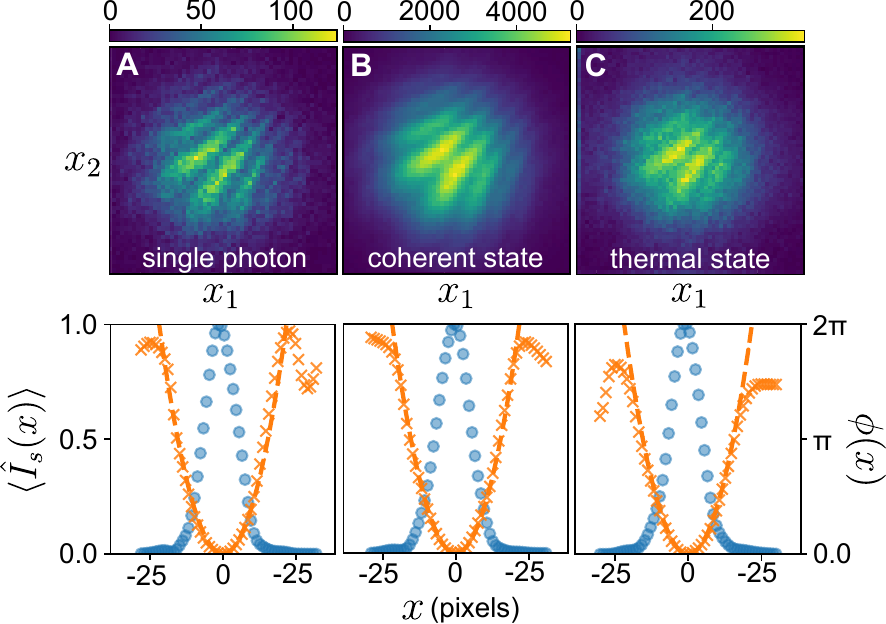}
\caption{\textbf{One-dimensional off-axis holography}.
(A) Heralded single photon. Event rate is 13 threefolds/s with $\tau_w = 5$~ns. 
(B) Coherent state. Event rate is $10^6$ twofolds/s with $\tau_w = 50~\mu$s. 
(C) Thermal state. Event rate is 5 twofolds/s with $\tau_w = 5$~ns. 
Top row: intensity correlation function $G(x_1,x_2)$ obtained with a quadratic phase pattern $\phi(x) = ax^2$ and a spatial shear between the signal and reference.
Colorbar shows the number of recorded correlation events at each pixel.
Bottom row: intensity $\braket{\hat{I}_s(x)}$ (blue dots, left axis) and phase $\phi(x)$ (orange crosses, right axis) of the signal recovered by a Fourier filtering procedure.
Dashed orange line is the expected phase applied by the SLM.
}
\label{fig:1d_phase}
\end{figure}

\subsection*{One-dimensional phase retrieval}
We start by considering a one-dimensional example.
We apply a quadratic phase pattern $\phi(x)=ax^2$ to the signal.
In addition, we tilt the signal beam along the $x$ direction with respect to the reference in order to introduce a shear, i.e. $\tilde{\phi}(x) =\phi(x) + k_0 x$ where $k_0$ is the strength of the shear. 
The measured intensity correlations for three different types of signals are shown in Fig.~\ref{fig:1d_phase}.
Since the phase is symmetric along the y-axis, we plot $G(x_1, x_2) = \sum_{y_1,y_2} G(x_1, y_1, x_2, y_2)$.
This quantity displays a fringe pattern along the anti-diagonal whose period is given by $2\pi/k_0$ and is slightly curved due to the quadratic phase. 
The fringe visibility $V$ is obtained by fitting the measured holograms to Eq.~\ref{eqn:intensity_corr_purestate}.
For each signal, we adjust the input intensity imbalance $\epsilon$ to maximize $V$ by rotating HWP1.
With $g^{(2)}_r = 1.006(11)$, the theoretical visibility upper limits are $82(1)\%$, $49.8(3)\%$, $41.8(1)\%$ for the single photon [$g^{(2)}_s = 0.049(8)$], coherent [$g^{(2)}_s = 1.006(11)$], and thermal [$g^{(2)}_s = 1.920(6)$] signal, respectively.
We observe $26.6(4)\%$, $14.0(1)\%$, and $11.3(3)\%$, which is roughly consistent with scaling the ideal visibilities $V_0$ by $M=0.3$.
The factor $M$ takes into account the various imperfections in our setup such as the temporal-spatial resolution of the camera and the mode overlap between the signal and reference (see Supplementary Materials for more information).

Although the visibilities are low, this has little impact on our ability to retrieve the signal phase.
We first demonstrate this using a Fourier filtering technique.
By taking a two-dimensional Fourier transform of the holograms in Fig.~\ref{fig:1d_phase}, the interference term is offset by $k_0$ from the background terms.
We isolate the interference term using a Gaussian window function and take the inverse Fourier transform to recover $\phi(x)$~\cite{schnars1994direct}.
This quantity is plotted alongside the directly measured intensity distribution of the signal $\braket{\hat{I}_s(x)}$ in the bottom row of Fig.~\ref{fig:1d_phase}.
We fit the recovered $\phi(x)$ to a quadratic function $ax^2$ and find the values $a=0.0135(3)$, $0.0133(1)$, and $0.0130(1)$ [rad/pixels$^2$] for the single photon, coherent, and thermal states, respectively, which are in agreement with the value applied to the SLM, $a=0.0132$.
The average pairwise fidelity between the three recovered complex-valued spatial modes is 0.98(1).

\begin{figure}
\centering
\includegraphics[width=1\columnwidth]{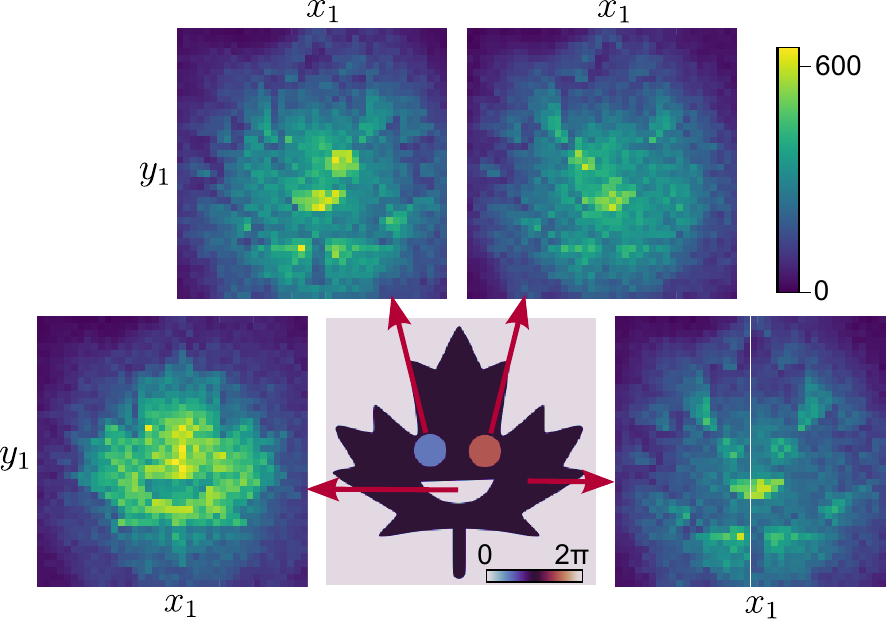}
\caption{
\textbf{Two-dimensional holograms.}
Center image shows the phase mask $\phi(\bm{r})$ applied to a coherent state signal.
Four surrounding images $I(\bm{r}_1) = I(x_1, y_1)$ are cross-sections of the measured intensity correlation function $G(\bm{r}_1, \bm{r}_2)$ at regions $\bm{r}_2$ indicated by the arrows.
From left to right, the phases in these regions are: (i) $\phi(\bm{r}_2)=0$, (ii) $\phi(\bm{r}_2)=\pi/2$, (iii) $\phi(\bm{r}_2)=3\pi/2$, (iv) $\phi(\bm{r}_2)=\pi$.
Colorbar shows number of recorded correlation events in 10~s.
}
\label{fig:cross_sections}
\end{figure}

\subsection*{Two-dimensional phase retrieval}
For a two-dimensional phase $\phi(\bm{r})$, it is no longer straightforward to use Fourier filtering since the intensity correlation function $G(\bm{r}_1, \bm{r}_2)$ is a four-dimensional quantity.
Thus, we turn to another phase-retrieval method.
According to Eq.~\ref{eqn:intensity_corr_purestate}, each cross-section $G(\bm{r}_1, \bm{r}_2)$ at $\bm{r}_2$ is an interference pattern $I(\bm{r}_1) \propto A-\cos{[\phi(\bm{r}_1)]}$ with a phase offset given by $\phi(\bm{r}_2)$.
As an example, we plot such cross-sections for a coherent state signal and the maple leaf phase mask shown in Fig.~\ref{fig:cross_sections}.
They can be interpreted as the left beam intensity images $I(\bm{r}_1)$ conditioned on detecting a right beam photon at $\bm{r}_2$, which provides the phase reference $\phi(\bm{r}_2)$.
Whenever $\phi(\bm{r}_1) = \phi(\bm{r}_2) + 2m\pi$, we observe destructive interference, whereas when $\phi(\bm{r}_1) = \phi(\bm{r}_2) + 2(m+1)\pi$, we observe constructive interference ($m \in \mathbb{Z}$).

The cross-sections of $G(\bm{r}_1, \bm{r}_2)$ are holograms which can be used to retrieve  $\phi(\bm{r})$.
However, we cannot use conventional phase-retrieval methods such as phase-shift holography since these assume knowledge of the offsets $\phi(\bm{r}_2)$~\cite{yamaguchi1993phase}.
Instead, we solve this self-referencing problem using a principal component analysis.
We construct a correlation matrix of all the holograms contained in $G(\bm{r}_1, \bm{r}_2)$:
\begin{equation}
\begin{split}
\Gamma(\bm{r}, \bm{r}') = \sum_{\bm{r}_2} G(\bm{r}, \bm{r}_2)G(\bm{r}', \bm{r}_2).
\end{split}
\label{eqn:correlation_matrix}
\end{equation}
This matrix describes which pixels in the holograms are in-phase (i.e. correlated) or out-of-phase (i.e. anti-correlated).
Because we are simply summing these correlations for all $\bm{r}_2$, $\Gamma(\bm{r}, \bm{r}')$ contains two orthogonal quadrature components varying with the sine and cosine of the phase mask wherever $\phi(\bm{r}_2)$ has values around $(m+1/2)\pi$ and $m\pi$, respectively.
We can isolate these components by performing a singular value decomposition of $\Gamma(\bm{r}, \bm{r}')$~\cite{vargas2011phase}.
This allows us to retrieve $\phi(\bm{r})$ without scanning the reference phase to specific values as in phase-shift holography.
Further details are given in the Methods.

\begin{figure}
\centering
\includegraphics[width=1\columnwidth]{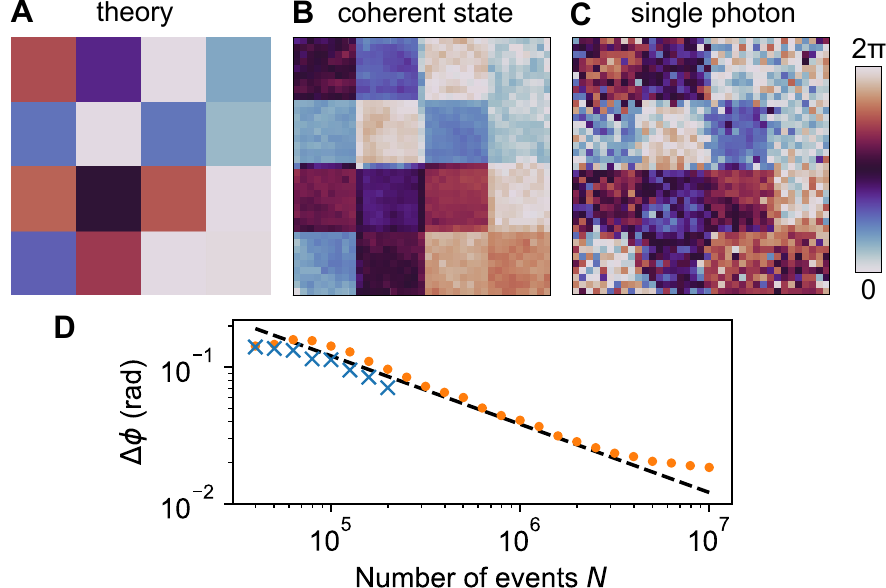}
\caption{
\textbf{Two-dimensional phase retrieval.}
(A) Checkerboard phase mask applied to SLM.
(B) Retrieved phase with a coherent state signal. Number of correlation events $N=1\times 10^7$. Acquisition time 10~s.
(C) Retrieved phase with a single photon signal. $N=2\times 10^5$. Acquisition time 10 hours.
(D) Precision of the retrieved phase for the coherent state (orange dots) and single photon (blue crosses).
The black dashed line is a fit to shot-noise scaling $\sim \sqrt{N}$.
}
\label{fig:2d_phase}
\end{figure}

We test our phase-retrieval method with the checkerboard phase pattern shown in Fig.~\ref{fig:2d_phase}A.
Using a coherent and single photon signal, we recover the phase shown in Fig.~\ref{fig:2d_phase}B and C, respectively, without background subtraction or optimization routines.
To quantify the precision of our technique, we compute the standard deviation of the recovered phase values in each checkerboard square, i.e. where the phase is expected to be uniform.
In Fig.~\ref{fig:2d_phase}D, we plot this phase uncertainty $\Delta \phi$, averaged over the center four squares where the beam intensity is largest.
For an equal number of correlation events, the single photon [blue crosses] shows slightly better performance than the coherent state [orange dots] due to the higher interference visibility of the former.
This performance follows shot-noise scaling $\sqrt{N}$ until $\sim 0.02$~rad, at which point the precision begins to plateau due to limitations of the SLM such as digitization and phase ripple errors.

\section*{Discussion}
Although our interferometer coherence time is on the order of 100~$\mu$s, we observe interference in our holograms despite using far longer integration times, e.g. 10~s with the coherent state signal, or 10~hours with the single photon signal. 
This demonstrates the first key advantage of using intensity interferometry for holography.
Namely, it is not necessary for the measured photons to be coherent over the entire integration time of the hologram.
This enables holography with extremely faint light and without active phase stabilization.
Some potential applications include quantitative phase imaging of biological samples which are susceptible to photodamage~\cite{park2018quantitative}, or performing standoff holography using a faint illumination beam.
Importantly, the phase-noise resilience of intensity interferometry does not require light with nonclassical correlations, which are often lost in practice due to attenuation and noise~\cite{tenne2019super,triginer2020quantum,casacio2021quantum}.

Our single photon and thermal state results demonstrate the second key advantage of intensity interferometry for holography.
Namely, it is possible to generate holograms of a signal beam using a reference derived from an independent light source.
This required mode-matching the two sources in polarization and time-frequency, which is generally quite challenging.
However, there are numerous applications when mode-matching can be achieved. 
For example, a single photon reference can potentially vastly extend the baseline of telescope arrays for astronomy~\cite{gottesman2012longer,brown2022interferometric}.   
Moreover, in quantum optics, it is practical to use a laser to characterize the mode structure of nonclassical states of light.
This idea has already been used to measure the time-frequency mode of photon pairs generated via spontaneous parametric down-conversion~\cite{thekkadath2022measuring}, and it could also be applied to photons produced by quantum dots~\cite{bennett2009interference} or single atoms~\cite{specht2009phase}.
With continuing improvements in the timing resolution of cameras, one could also envisage tracking the interference between two independent lasers which are not phase-locked nor frequency-locked~\cite{magyar1963interference,pfleegor1967interference,liu2015second}.

As a final remark, we emphasize that intensity interference only requires the ability to measure field intensities.
For faint light sources which produce only a few photons per coherence time, such as the sources considered in this work, these correlations must be measured using single-photon detectors.
However, for brighter sources, they could also be measured using a regular camera, as long as each frame is captured with a sufficiently short exposure time (see Correlation Window section in Methods). 
In this case, $G(\bm{r}_1,\bm{r}_2)$ would simply be determined by the product of the intensities measured by pixels $\bm{r}_1$ and $\bm{r}_2$, averaged over many frames.
Bright sources also naturally exhibit intensity correlations beyond second-order, which, when resolved, can improve the contrast of images~\cite{agafonov2008high}.

\textit{Note}: During the preparation of this manuscript, we became aware of a related work~\cite{Szuniewicz2023noise}.

\section*{Methods}

\subsection{Signal and reference preparation}
We describe the setup shown in Fig.~\ref{fig:setup}A in more detail.
A titanium sapphire oscillator (Coherent Chameleon Ultra II) generates an 80 MHz train of 150 fs pulses with a center wavelength of 810 nm. 
The pulses are focused into a 1-mm-long beta barium borate (BBO) crystal for second harmonic generation.
The frequency-doubled light (approximately 1 W) is separated using a dichroic mirror and is used to pump a type-II spontaneous parametric down-conversion process in a 2-mm-long BBO crystal.
The degenerate photon pairs are split on a polarizing beam splitter (PBS).
The horizontally-polarized photon is coupled directly into a single-mode fiber and is detected by an avalanche photodiode (APD, Excelitas SPCM-AQRH).
We detect roughly $3\times10^5$ photons per second in this heralding arm.
The vertically-polarized photon is overlapped with attenuated light from the pump laser on a PBS.
Both are coupled into a polarization-maintaining fiber which leads to the interferometer part of the experiment.
All modes are spectrally filtered by a bandpass filter (3 nm full-width-half-max, Semrock LL01-810).

Using this setup, we can prepare the signal in three different types of states: (i) heralded single photon, (ii) thermal state, (iii) phase-randomized coherent state.
In case (i), we send the APD electronic signal to the camera time-to-digital converter and include the condition of detecting a herald photon within the correlation window, i.e. measure threefold coincidences.
The time delay $\tau_1$ is adjusted so that the single photon and coherent state arrive at the PBD simultaneously.
In case (ii), we ignore the herald photon and thus the signal mode produced by the down-conversion source has thermal statistics.
Finally, in case (iii), we insert a beam block (B) and only use the attenuated coherent state.
We adjust HWP2 to split the light between both arms of the interferometer.
The time delay $\tau_2$ is adjusted so that both interferometer paths are matched in length.

In all three cases, the spatial light modulator (Meadowlarks E-Series 1920 $\times$ 1200) is mounted on a vibrating piezo-stage (Thorlabs NF15AP25) driven by a voltage source (Thorlabs MDT693A) producing a 75 V peak-to-peak and 100 Hz frequency sine curve.
The piezo vibrations reduce the interferometer coherence time to $\sim 100~\mu$s which ensures that no first-order interference is visible over our integration times.

\subsection{Theory}
We derive the intensity cross-correlation function at the output of the beam splitter.
We assume the input fields are only phase-stable for a finite coherence time $\tau_c$.
For times longer than $\tau_c$, we model the phase incoherence as a randomly fluctuating phase difference $\theta(t)$.
This phase has no effect on rotationally-symmetric quantum states such as a single photon or thermal state, but we include it in our derivation for the sake of generality.
The instantaneous intensities at the output of the beam splitter are given by $\hat{I}_{c,d}(\bm{r}; t) = \hat{E}^-_{c,d}(\bm{r}; t)\hat{E}^+_{c,d}(\bm{r}; t)$, where the subscripts $c$ and $d$ label the output mode.
The detectors record these intensities averaged over some response time $\tau_d$.
We assume that $\tau_d$ is longer than any dynamics in the fields (e.g. the pulse duration) but is short enough that $\theta(t)$ can be considered as constant, i.e. $\theta(t) \equiv \theta$.
This allows us to drop the time coordinates:
\begin{align}
\begin{split}
\hat{I}_{c,d}(\bm{r}) &= \int_{-\tau_d/2}^{\tau_d/2} dt \hat{I}_{c,d}(\bm{r}; t) \\
&= \frac{1}{2}(\hat{I}_s(\bm{r})+\hat{I}_r(\bm{r}) \\
&\qquad \pm[\hat{E}^-_s(\bm{r})\hat{E}_r^+(\bm{r})e^{i\theta}+c.c.]),
\end{split}
\label{eqn:integrated_intensity}
\end{align}
where the $+$ ($-$) corresponds to the intensity in mode $c$ ($d$). 
We used the beam splitter transformation to express the output intensities in terms of the input fields.
A derivation which does not drop the time coordinate and describes the effect of detector timing resolution and temporal-spectral mode mismatch is given in Ref.~\cite{ou1988quantum}.
If we choose $\tau_d$ as the correlation window, the intensity correlations measured by the detectors in a single trial are given by $\hat{I}_{c}(\bm{r}_1)\hat{I}_{d}(\bm{r}_2)$ with the phase $\theta$ staying constant in each trial.
However, over many trials, $\theta$ fluctuates randomly between $[0, 2\pi]$.
Thus the measured intensity correlation averaged over all trials is given by the classical ensemble average, which we denote using the brackets $\{ \}$:
\begin{equation}
\{ \hat{I}_{c}(\bm{r}_1)\hat{I}_{d}(\bm{r}_2) \} = \int_0^{2\pi} d\theta \hat{I}_{c}(\bm{r}_1)\hat{I}_{d}(\bm{r}_2).
\label{eqn:classical_ensemble_avg}
\end{equation}
Expanding Eq.~\ref{eqn:classical_ensemble_avg} in terms of the input field operators [Eq.~\ref{eqn:integrated_intensity}] reveals 16 terms. 
Those which contain an odd number of creation or annihilation operators in either input mode vanish due to the ensemble average, and we are left with only six terms:
\begin{equation}
\begin{split}
\{ \hat{I}_{c}(\bm{r}_1)\hat{I}_{d}(\bm{r}_2) \}
&= \frac{1}{4}[\hat{I}_s(\bm{r}_1)\hat{I}_r(\bm{r}_2) + \hat{I}_s(\bm{r}_2)\hat{I}_r(\bm{r}_1) \\ 
&\quad + \hat{E}_s^-(\bm{r}_1) \hat{E}_s^-(\bm{r}_2) \hat{E}_s^+(\bm{r}_1) \hat{E}_s^+(\bm{r}_2) \\ 
&\quad + \hat{E}_r^-(\bm{r}_1) \hat{E}_r^-(\bm{r}_2) \hat{E}_r^+(\bm{r}_1) \hat{E}_r^+(\bm{r}_2) \\
&\quad - \hat{E}_r^-(\bm{r}_1) \hat{E}_r^+(\bm{r}_2)\hat{E}_s^-(\bm{r}_2) \hat{E}_s^+(\bm{r}_1)  \\
&\quad - \hat{E}_r^-(\bm{r}_2) \hat{E}_r^+(\bm{r}_1)\hat{E}_s^-(\bm{r}_1) \hat{E}_s^+(\bm{r}_2)] \\
\end{split}
\label{eqn:averaged_intensity_correlations}
\end{equation}
The last two terms correspond to the interference between the input fields.
The measured intensity correlations are given by the expectation value of  Eq.~\ref{eqn:averaged_intensity_correlations} with respect to the input quantum states, i.e. $G(\bm{r}_1, \bm{r}_2) = \braket{\{ \hat{I}_{c}(\bm{r}_1)\hat{I}_{d}(\bm{r}_2) \}}$.
We divide this quantity by the average input intensities (and assume $\epsilon = \braket{\hat{I}_r(\bm{r})}/\braket{\hat{I}_s(\bm{r})}$ is independent of $\bm{r}$) to obtain:
\begin{equation}
\begin{split}
\tilde{G}(\bm{r}_1, \bm{r}_2) &= \frac{ \braket{\{ \hat{I}_{c}(\bm{r}_1)\hat{I}_{d}(\bm{r}_2) \}} } {\braket{\hat{I}_s(\bm{r}_1)}\braket{\hat{I}_r(\bm{r}_2)}} \\
&= \frac{1}{4}( 2 + \frac{1}{\epsilon} g_s^{(2)}(\bm{r}_1,\bm{r}_2) + \epsilon g_r^{(2)}(\bm{r}_1,\bm{r}_2) \\
&\qquad - 2\mathrm{Re}[g_s^{(1)}(\bm{r}_1,\bm{r}_2)g_r^{(1)}(\bm{r}_1,\bm{r}_2)^*] ),
\end{split}
\label{eqn:intensity_correlations_coherence_functions}
\end{equation}
such that Eq.~\ref{eqn:intensity_correlations_coherence_functions} is now cast in terms of the normalized coherence functions,
\begin{align}
g_{s,r}^{(1)}(\bm{r}_1,\bm{r}_2) &= \frac{\braket{\hat{E}_{s,r}^-(\bm{r}_1) \hat{E}_{s,r}^+(\bm{r}_2)}}{[\braket{\hat{I}_{s,r}(\bm{r}_1)}\braket{\hat{I}_{s,r}(\bm{r}_2)}]^{1/2}}, \\
g_{s,r}^{(2)}(\bm{r}_1,\bm{r}_2) &= \frac{\braket{\hat{E}_{s,r}^-(\bm{r}_1)\hat{E}_{s,r}^-(\bm{r}_2)\hat{E}_{s,r}^+(\bm{r}_1)\hat{E}_{s,r}^+(\bm{r}_2)}}{\braket{\hat{I}_{s,r}(\bm{r}_1)}\braket{\hat{I}_{s,r}(\bm{r}_2)}}.
\end{align}
Eq.~\ref{eqn:intensity_correlations_coherence_functions} is 
shown as Eq.~\ref{eqn:intensity_cross_correlation} in the main text.

\subsection{Correlation window}
The role of the correlation window warrants further discussion.
We can distinguish between two scenarios.
In the first, we consider correlated input fields $\hat{E}_s$ and $\hat{E}_r$ such as photon pairs produced by parametric down-conversion.
Then, $\braket{\hat{I}_s(\bm{r}_1)\hat{I}_r(\bm{r}_2)} \neq \braket{\hat{I}_s(\bm{r}_1)}\braket{\hat{I}_r(\bm{r}_2)}$ and intensity interference can be observed even when the correlation window (and thus the detection time resolution) is significantly longer than the coherence time of the input fields~\cite{ou1988quantum}.
This effect is well-known in Hong-Ou-Mandel interference~\cite{hong1987measurement}.
In a related scenario, one splits a single light source at a beam splitter in order to obtain the fields $\hat{E}_s$ and $\hat{E}_r$, i.e. these are inside an interferometer with balanced arm lengths.
In this case, the relevant coherence time is not that of the source, but rather the time over which the arms of the interferometer are phase stable~\cite{ou1989fourth}.
We encounter this scenario in our experiment with the phase-randomized coherent state.
With the piezo-stage vibrating, the interferometer arms are stable over roughly $100~\mu$s.
In practice, we use a correlation window of $\tau_w = 50~\mu$s.
Thus, the data acquisition rate is greatly increased by capturing correlation events obtained between many consecutive pulses~\cite{kim2013conditions}.

In a second scenario, we consider that the signal and reference originate from independent light sources.
We encounter this scenario in our experiment with the heralded single photon and thermal signal.
To observe intensity interference, the correlation window should generally be chosen to be shorter than the coherence time of the input fields~\cite{mandel1983photon}.
One exception is if non-stationary light fields (e.g. pulses) are employed, as in our experiment. 
Although our pulses have picosecond coherence times, we still observe interference with a $\tau_w = 5$~ns correlation window.
Because $\tau_w$ is shorter than the inter-pulse separation of 12.5 ns and light is absent between pulses, a detector click could have only occurred from a photon within the coherence time (or a dark count).

\subsection{Camera}
We use a complementary metal–oxide–semiconductor event camera (Amsterdam Scientific Instruments TPX3CAM).
It has a resolution of 256 $\times$ 256 pixels with a pixel pitch of 55 $\mu$m.
To achieve single-photon sensitivity, an intensifier (Photonis Cricket) is placed before the camera~\cite{nomerotski2019imaging}.
This first converts incident photons into electrons which are then amplified by a multichannel plate before finally striking a phosphor screen.
The flash produced by the intensifier is detected by the camera as a cluster of pixels localized in space and time.
The detection occurs with a total quantum efficiency of 7(2)\%~\cite{vidyapin2023characterisation}.
We process the raw camera data using a real-time clustering algorithm to produce a stream of the pixel coordinate and arrival time of the detected photons.
For each pulse of light from the laser, the full width at half maximum of the timestamp distribution is $\tau_d = 8.3(6)$~ns due to the detector timing resolution.
We operate the camera far from a saturation regime such that the likelihood of there being two or more photons at the same pixel within its dead time ($\sim 1~\mu$s) is negligible.

\subsection{Principal component phase retrieval}
The method of Ref.~\cite{vargas2011phase} can be adapted to work with intensity interferometry data.
Each cross-section of the intensity correlation function $\braket{\hat{I}(\bm{r}_1)\hat{I}(\bm{r}_2)}$ at a particular $\bm{r}_2=\bm{r}_n$ is a hologram obtained with the phase reference value $\phi(\bm{r}_2) \equiv \delta_n$.
Its intensity is given by
\begin{equation}
\begin{split}
I_n(\bm{r}) &= B_n(\bm{r}) - \braket{\hat{I}_s(\bm{r})}\braket{\hat{I}_r(\bm{r}_n)}\cos{[\phi(\bm{r})-\delta_n]}/2 \\ 
&= B_n(\bm{r}) - \alpha_n I_{\mathrm{sin}}(\bm{r})  - \beta_n I_{\mathrm{cos}}(\bm{r}),
\end{split}
\label{eqn:hologram_cos_sin}
\end{equation}
where $B_n(\bm{r})= \braket{\hat{I}_s(\bm{r})}\braket{\hat{I}_r(\bm{r}_n)}(  2 + g_s^{(2)}/\epsilon + \epsilon g_r^{(2)})/4$ is a background term.
In the second line, we used a trigonometric identity to express the equation in terms of a sine and cosine term,
\begin{align}
I_{\mathrm{sin}}(\bm{r}) &=\braket{\hat{I}_s(\bm{r})}\sin{[\phi(\bm{r})]} \\
I_{\mathrm{cos}}(\bm{r}) &=\braket{\hat{I}_s(\bm{r})}\cos{[\phi(\bm{r})]}
\end{align}
with coefficients $\alpha_n = \braket{\hat{I}_r(\bm{r}_n)}\sin{(\delta_n)}/2$ and $\beta_n = \braket{\hat{I}_r(\bm{r}_n)}\cos{(\delta_n)}/2$, respectively.
These two terms are orthogonal components of the intensity in the sense that they are uncorrelated, i.e.
\begin{equation}
\sum_{\bm{r}} I_{\mathrm{sin}}(\bm{r})I_{\mathrm{cos}}(\bm{r}) \approx 0.
\end{equation}
Thus, we can isolate them with a principal component analysis.
We construct a correlation matrix for all $n$ (i.e. $\bm{r}_2$):
\begin{equation}
\Gamma(\bm{r}, \bm{r}') = \sum_{n} I_n(\bm{r})I_n(\bm{r}').
\end{equation}
The principal components of $\Gamma(\bm{r}, \bm{r}') \equiv \Gamma$ are found by performing a singular value decomposition:
\begin{equation}
\Gamma = U \Sigma V^*.   
\end{equation}
The first column of $U$ is the largest principal component of $\Gamma$ and relates to the background term.
The next two columns respectively yield $I_{\mathrm{sin}}(\bm{r})$ and $I_{\mathrm{cos}}(\bm{r})$.
We recover the phase using
\begin{equation}
\phi(\bm{r}) = \mathrm{arctan}\left( \frac{I_{\mathrm{sin}}(\bm{r})}{I_{\mathrm{cos}}(\bm{r})} \right).
\end{equation}

\section{Acknowledgements}
We thank Rune Lausten and Denis Guay for their technical support.
We also thank Andrei Nomerotski and Jingming Long for their help with the camera.
This work was funded by: Natural Sciences and Engineering Research Council of Canada; Engineering and Physical Sciences Research Council (T00097X and P510257); Korea Institute of Science and Technology open research programme.

\section{Author contributions}
G.S.T and D.E. developed the concept.
G.S.T. performed the experiment with assistance from D.E., F.B., and Y.Z., while M.S.K and B.S. supervised the work.
All authors contributed to writing the manuscript.

\bibliographystyle{apsrev4-2}
\bibliography{refs}

\newpage
\setcounter{figure}{0}
\setcounter{equation}{0}
\renewcommand{\theequation}{S\arabic{equation}}
\renewcommand{\thefigure}{S\arabic{figure}}

\section*{Supplemental Material}

\subsection*{Interference visibility}

\begin{figure}
\centering
\includegraphics[width=0.70\columnwidth]{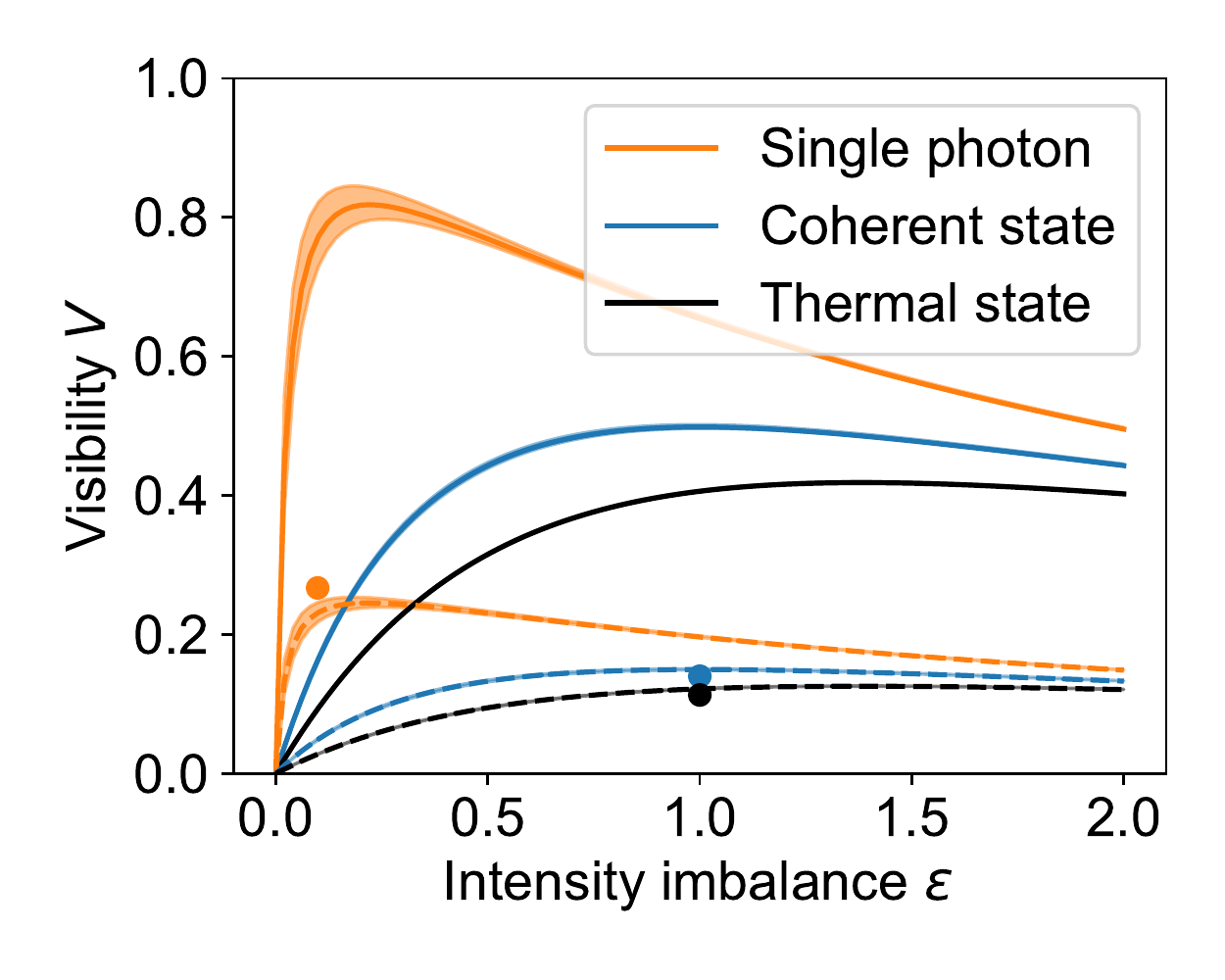}
\caption{\textbf{Visibility}. 
Visibility $V$ [Eq.~\ref{eqn:modified_vis}] of intensity interference as a function of the average input intensity imbalance $\epsilon=\braket{\hat{I}_{r}}/\braket{\hat{I}_{s}}$.
Continuous lines use $M=1$ while the dashed lines use $M=0.3$.
The line thickness shows one standard deviation of uncertainty due to the measurement error in $g^{(2)}_r$ and $g^{(2)}_s$.
Markers are experimental data points.
}
\label{fig:vis}
\end{figure}

In Eq.~\ref{eqn:intensity_corr_purestate}, we showed that the visibility of intensity interference depends on the photon-number statistics of the signal and reference, i.e. $V_0=2/A$ where $A=2+g^{(2)}_s/\epsilon + \epsilon g_r^{(2)}$.
That derivation assumed the interfering fields occupied the same temporal-spectral mode, had identical polarisation, and were measured by ideal detectors.
Rather than individually modeling each possible imperfection (e.g. see Ref.~\cite{ou1988quantum} for a treatment of temporal distinguishability), we can simply model their combined effect to the visibility by introducing a phenomenological parameter $M$:
\begin{equation}
V = MV_0 = \frac{2M}{2+g^{(2)}_s/\epsilon + \epsilon g_r^{(2)}}.
\label{eqn:modified_vis}
\end{equation}
We determine $M$ empirically using our observed interference visibilities.
In Fig.~\ref{fig:vis}, we plot Eq.~\ref{eqn:modified_vis} using the measured $g^{(2)}$ values (i.e. $g^{(2)}_r = 1.006(11)$ and $g^{(2)}_s = 0.049(8)$, $1.006(11)$, $1.920(6)$ for the single photon, coherent, and thermal signal, respectively).
We find reasonable agreement with our data points [markers] using $M=0.3$ and no other fitting parameters.
In the next section, we describe the contribution of various imperfections to $M$.

\subsection*{Imperfections contributing to $M$}
We present various measurement to give a rough idea of the relative contribution of each imperfection to the reduced visibility.
The results are summarized in Table 1.

\begin{table}
\centering
\begin{tabular}{|c  c|}
\hline
Imperfection                &  Estimated reduction to $V$ \\
\hline
Mode overlap                & 85\%  \\
Camera noise                & 70\%  \\
Camera temporal resolution  & 95\%  \\
Camera spatial resolution   & 50\%  \\
Total                       & $M\sim 30\%$ \\
\hline
\end{tabular}
\label{table:imperfections}
\caption{Estimate of the contribution of various imperfections to the reduced visibility $V$.}
\end{table}

\subsubsection*{Mode overlap and camera noise}
\begin{figure*}
\centering
\includegraphics[width=0.9\linewidth]{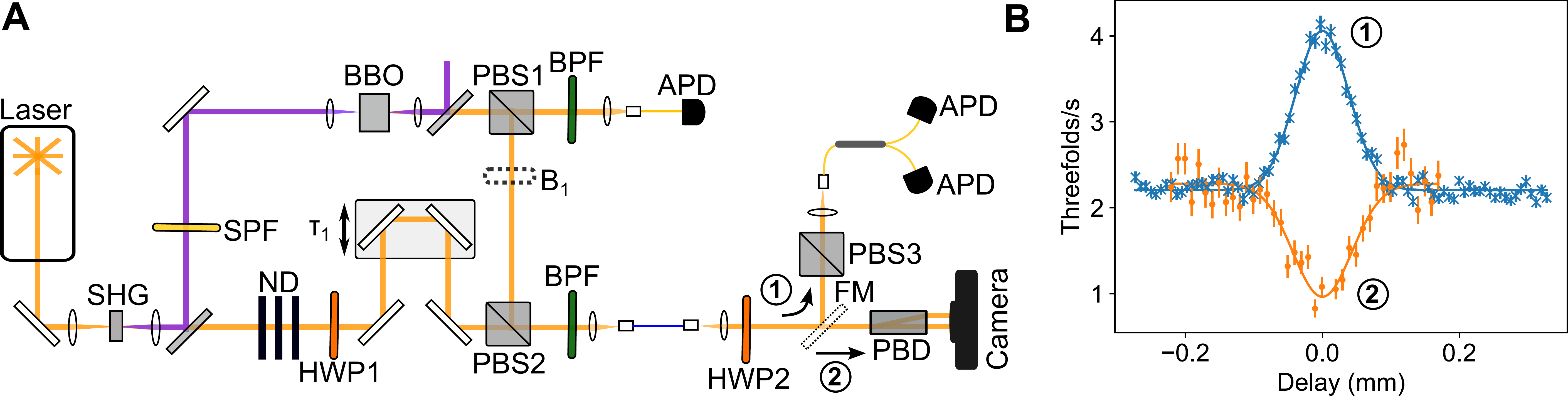}
\caption{\textbf{Hong-Ou-Mandel-type interference}. 
(A) Experimental setup used to characterize the interference visibility using APDs (path 1) or the camera (path 2).
The signal is a heralded single photon state, while the reference is a weak coherent state.
(B) Observed peak (dip) in the measured threefolds as a function of the delay $\tau_1$ measured by the APDs (camera).
The visibility of the peak (dip) is 84(2)\% (58(4)\%).
}
\label{fig:HOM}
\end{figure*}

We characterized the mode overlap (i.e. polarization and temporal-spectral)  between the heralded single photon signal and coherent state reference using a Hong-Ou-Mandel-type measurement.
The setup is shown in Fig.~\ref{fig:HOM}A.
Using a flip mirror (FM), we switched between measuring the interference using APDs (path 1) or the camera (path 2).
The threefolds measured with the APDs [blue crosses] are plotted in Fig.~\ref{fig:HOM}B as a function of the delay $\tau_1$.
We observe a peak rather than the conventional dip since we are detecting bunching at the output port of PBS3.
The peak has a visibility of $V = 84(2)\%$, where $V = (C_\mathrm{max}-C_\mathrm{min})/C_\mathrm{max}$ and $C_\mathrm{max}$ and $C_\mathrm{min}$ are the maximum and minimum count rate obtained from a Gaussian fit.
The observed visibility is smaller than the upper limit given by the $\sim$92\% spectral modal purity of the SPDC source inferred from the unheralded second-order autocorrelation measurement, $g^{(2)}=1.920(6)$~\cite{thekkadath2022measuring}.

We then switch to measuring this interference signal with the camera.
In this case, we treat the camera region at each output of the PBD as a bucket detector.
The threefolds [orange dots] are plotted in Fig.~\ref{fig:HOM}B and show a dip with a visibility of 58(4)\%.
We attribute the lower interference visibility measured by the camera to three main factors: (i) increased dark counts ($\sim 5\times10^4$ counts/s) (ii) reduced temporal resolution (see below), (iii) increased multiphoton noise.
This last effect is related to the lower detection efficiency of the camera ($\sim 7\%$) compared to the APDs ($\sim 60\%$): in order to achieve a similar threefold count rate when using the camera, we increased the SPDC pump power and reference intensity, which increases the multiphoton noise.

We note that the imperfect mode overlap should not contribute to $M$ for the coherent state signal, since in this case, both the signal and reference fields are derived from the same source.


\subsubsection*{Camera temporal resolution}
\begin{figure}
\centering
\includegraphics[width=0.70\columnwidth]{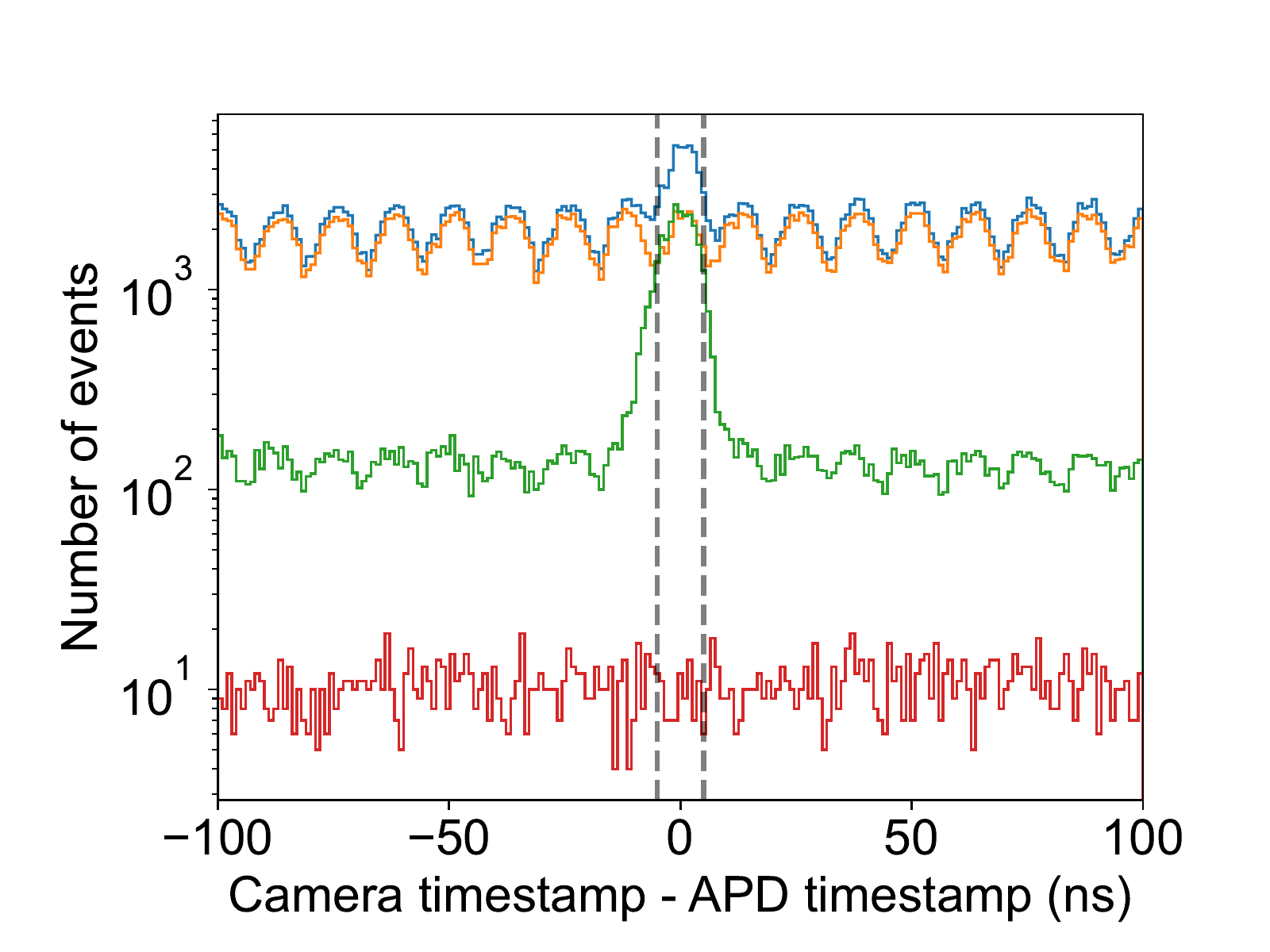}
\caption{\textbf{Timestamp histogram}. 
Difference between the timestamp produced by a herald photon arriving at the APD and a photon arriving at the time-tagging camera in 10~s of integration.
The signal is a heralded single photon and the reference is a coherent state.
Blue: both signal and reference unblocked, orange: signal blocked, green: reference blocked, red: both blocked.
Dashed grey lines show the correlation window $\tau_w = 5$ ns.
}
\label{fig:SNR}
\end{figure}

The finite temporal resolution of the camera causes accidental correlation events which also contribute to reducing $V$.
Their origin can be seen in Fig.~\ref{fig:SNR}.
The blue line is a histogram of the timestamp difference between a photon detected by the camera (in either the left or right beam region) and a herald photon detected by the APD.
One can observe a train of pulses separated by 12.5 ns (corresponding to the repetition rate of the laser) with a peak near $t=0$ due to the correlated photon pairs produced by SPDC.
While the pulses themselves are only picoseconds in duration, their shape is convolved by the camera response time $\tau_d\sim 8.3(6)$ns.
Thus, photons from different pulses have a chance to produce timestamps within the correlation window, which contribute to accidentals.
By fitting Gaussians to the first three peaks around $t=0$, we estimate that these accidentals correspond to roughly 2\% of the measured threefolds for a correlation window of size $\tau_w = 5$~ns.
In the same figure, we show histograms obtained with the heralded photon blocked (orange), the coherent state blocked (green), and both blocked (red).

\subsubsection*{Camera spatial resolution}
\begin{figure}
\centering
\includegraphics[width=0.70\columnwidth]{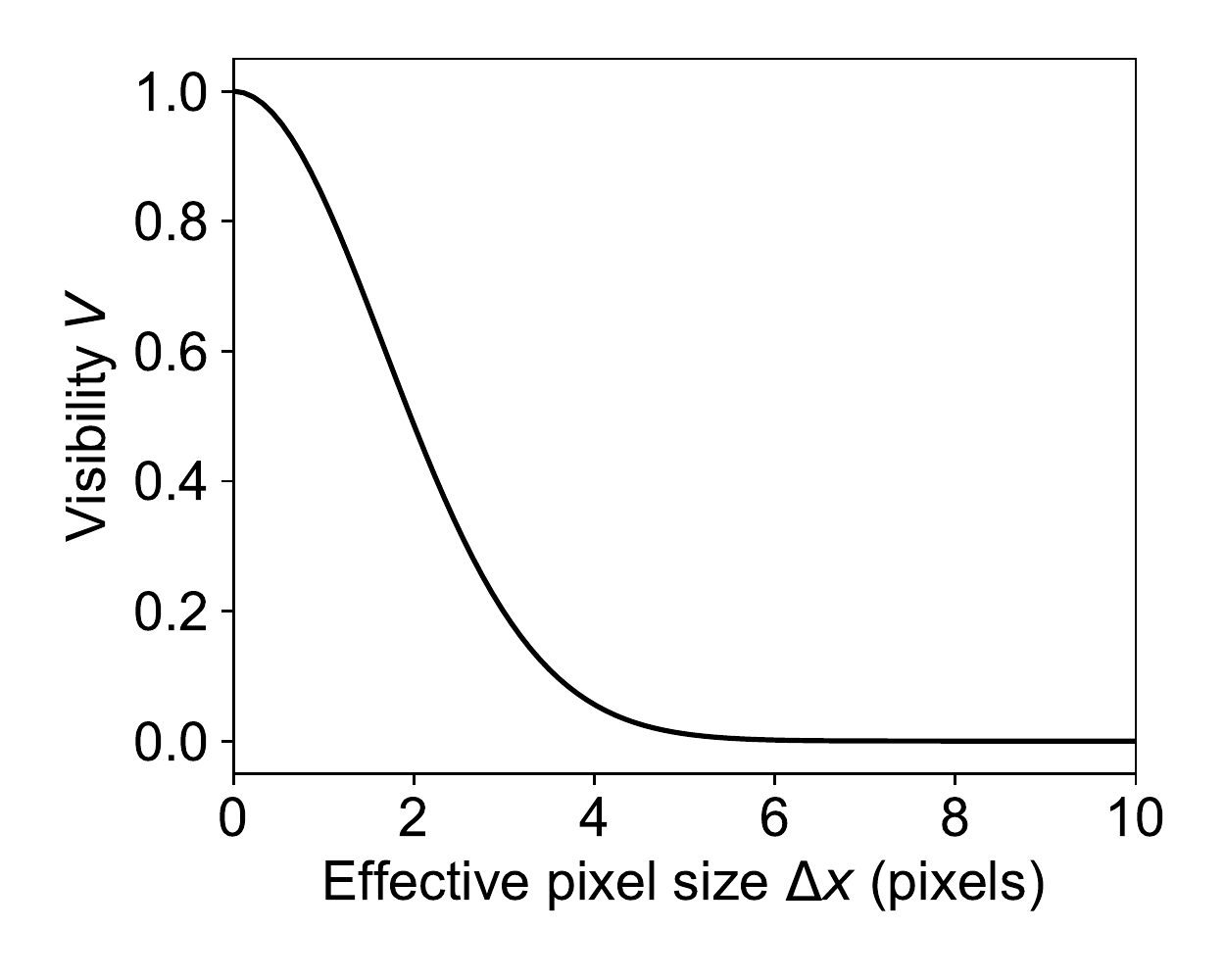}
\caption{\textbf{Visibility and spatial resolution}. 
The observed visibility $V$ of the intereference pattern diminishes as the effective pixel size of the camera $\Delta x$ increases. Here we assumed the interference fringe period is given by $2\pi/k_0$, where $k_0=0.62$ [1/pixel] is shear strength used in the experiment.}
\label{fig:vis_spatial_res}
\end{figure}

There remains a substantial discrepancy between the 58(4)\% visibility measured in Fig.~\ref{fig:HOM}(b) compared to the 26.6(4)\% measured in Fig.~\ref{fig:1d_phase}.
We attribute the remaining reduction in $V$ to the finite spatial resolution of the camera.
In the latter measurement, we introduced a shear between the signal and reference beams in order to use off-axis holography.
The shear had a strength of $k_0 = 0.62(2)$ [1/pixels] leading to an interference pattern with a fringe period of $2\pi/k_0\sim$ 10 pixels.
Due to the camera intensifier, a single photon causes a cluster of pixels to fire, typically containing 6-7 pixels~\cite{vidyapin2023characterisation}.
We assigned the pixel with the largest response (i.e. time-over-threshold) as the spatial coordinate for that detection event.
This leads to a blurring effect which reduces the effective resolution of the camera.
We model this effect by convolving the expected fringe pattern with a Gaussian of standard deviation $\Delta x$ [Fig.~\ref{fig:vis_spatial_res}].
An effective pixel size of $\Delta x \sim 2$ pixels accounts for the remaining discrepancy in the observed visibility.

\end{document}